\documentclass{article}
\usepackage{frascatiphys}
\usepackage{graphicx}
\begin{document}
\title{ 
HEAVY QUARKONIA
}
\author{
Zaza Metreveli        \\
{\em Department of Physics, Northwestern University, Evanston, IL 60208, USA} \\
}
\maketitle
\baselineskip=11.6pt
\begin{abstract}
Recent experimental results on heavy quarkonia spectroscopy and decays are 
reviewed. In particular, new results are discussed on charmonium spin singlet 
states, bottomonium D-states, photon and hadronic transitions from heavy 
quarkonium states, and the unexplained narrow X(3872) state. 
\end{abstract}
\baselineskip=14pt
\section{Introduction}
Heavy quarkonia are the bound states of charm and bottom quarks.
They are strong interaction analogs of positronium. Because
charm and bottom quarks have large masses ($\sim$1.5 and 
$\sim$4.5 GeV), velocities of quarks in hadrons are nonrelativistic. The
strong coupling constant $\alpha_{s}$ is small ($\sim$0.3 for $c\bar{c}$
and $\sim$0.2 for $b\bar{b}$). Therefore heavy quarkonia 
spectroscopy is a good testing ground for theories of strong interactions:
QCD in both perturbative and non-perturbative regimes, QCD inspired 
purely phenomenological potential models, NRQCD and Lattice QCD.
\par
\begin{figure}[t]
\vspace{6.0cm}
\includegraphics{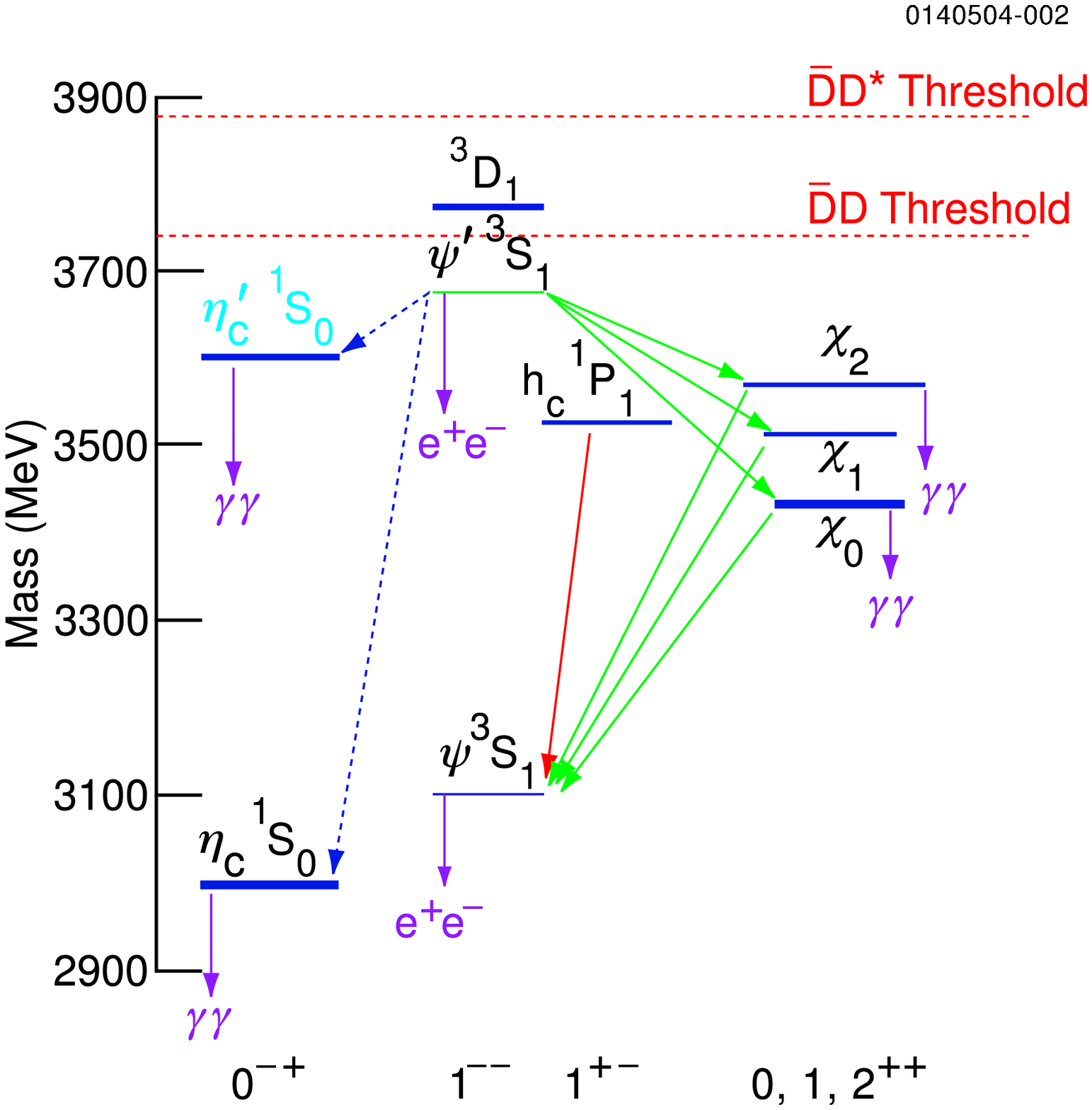}
\includegraphics{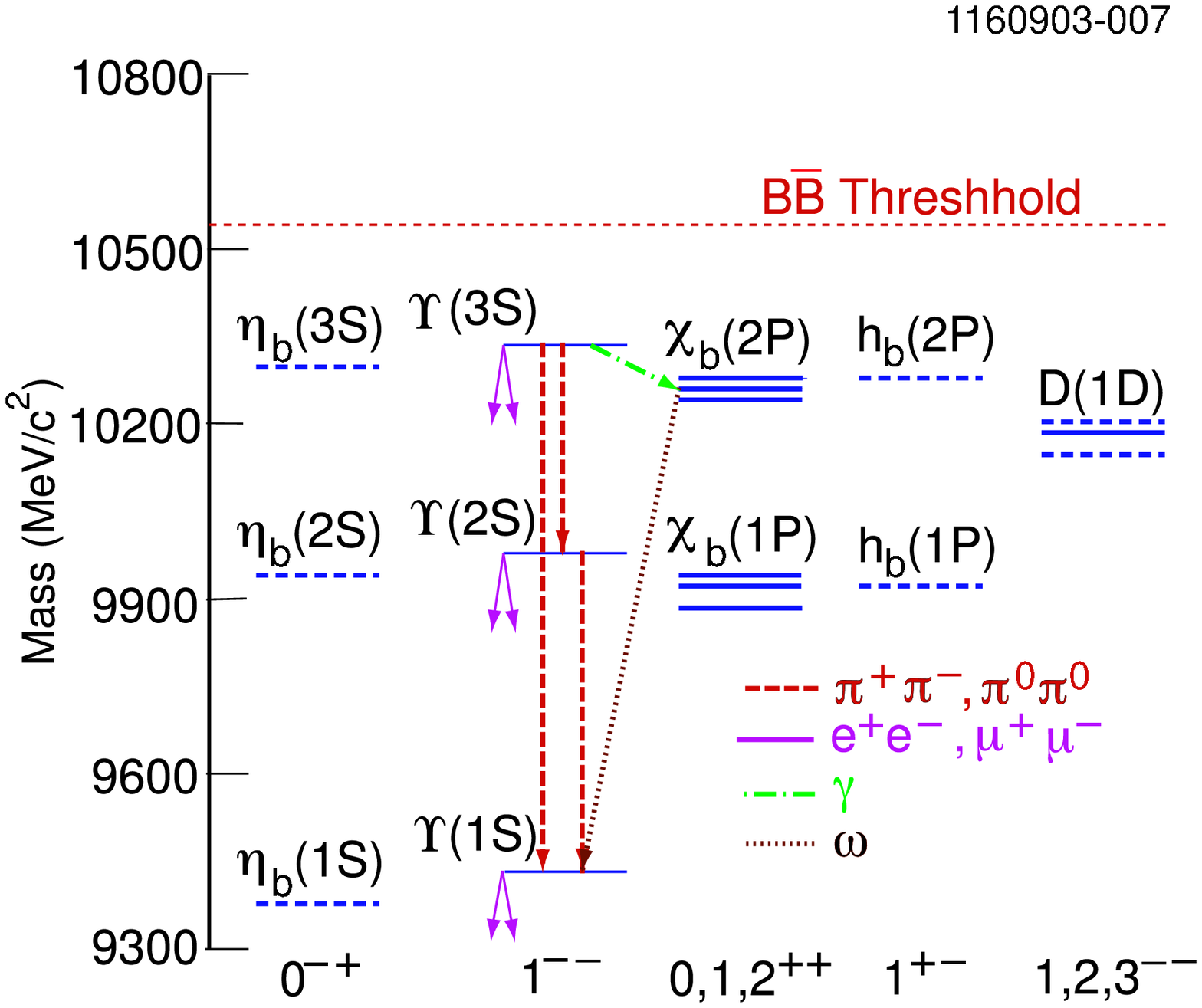}
 \caption{\it The spectra for charmonia (left) and bottomonia (right) below and near the open flavor threshold. Some typical transitions are indicated. None of the singlet $\eta_{b}$ \it or $h_{b}$ \it bottomonium states or $h_{c}$ \it charmonium state have been observed yet.\label{spectras} }
\end{figure}

Quarkonium states can be produced (fig.\ref{spectras}) in 
$e^{+}e^{-} \to \gamma^{*} \to q \bar{q}$ processes 
(direct production of $n^{3}S_{1}$: $J^{PC}=1^{--}$ vector states), 
two photon fusion processes at the $e^{+}e^{-}$ colliders 
(production of $\eta_{c}$, $\eta_{c}^{\prime}$, $\chi_{0,2}$: 
$J^{PC}=0^{-+},0^{++},2^{++}$ states), $p\bar{p}$ annihilation via 
two or three gluons (production of $q\bar{q}$ mesons with any quantum 
numbers), B meson decays (production of states with any quantum numbers
with associated particles), radiative or hadronic transitions from higher 
states of quarkonia. 
\section{New in Charmonium Spectroscopy} 
In this section new experimental results on charmonium states 
produced below the open flavor production threshold are reviewed,
obtained from the large data samples collected with the BaBar, Belle, 
BES and CLEO detectors.
\subsection{Spin Singlet States}
These states are generally the most difficult states to access and study
because they are not directly formed in $e^{+}e^{-}$ annihilation and  
the radiative decays of spin triplet states $^{3}S_{1}$($J/\psi$, 
$\psi^{\prime}$, $\Upsilon$, ...) to spin singlet states 
$^{1}S_{0}(\eta_{c},\eta_{b})$ are M1 transitions and therefore are highly 
suppressed. In $^{1}P_{1}(h_{c},h_{b})$ cases, the radiative decays are 
entirely forbidden by C-conservation. As a result, no singlet states have 
ever been identified in bottomonium and only the $\eta_{c}$ singlet state 
was identified in charmonium until recently.
\begin{figure}[t]
\vspace{6.0cm}
\includegraphics{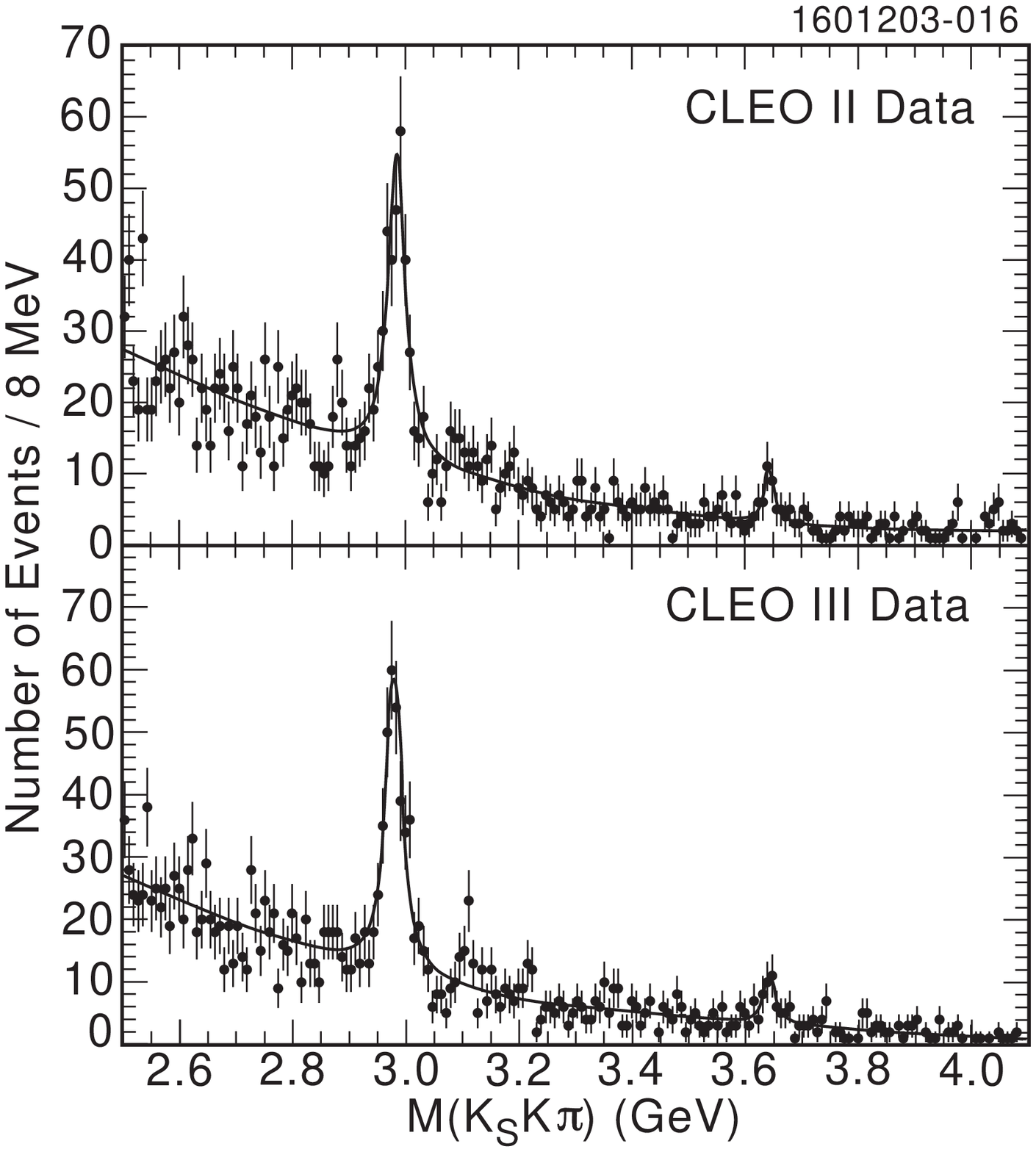}
\includegraphics{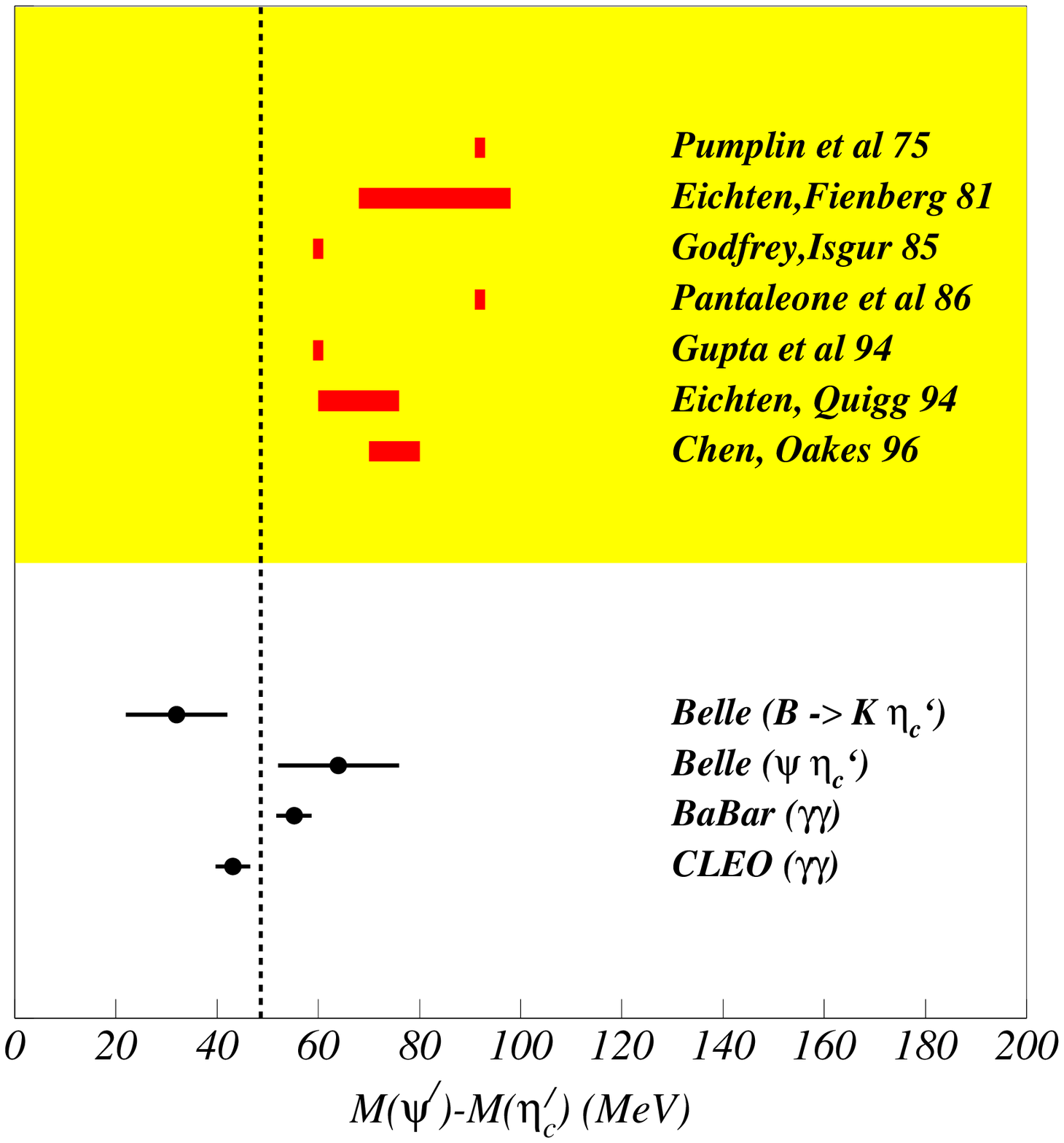}
\caption{\it $K_{s}K^{\pm}\pi^{\mp}$ \it invariant mass distributions for events in two photon fusion processes from the CLEO data, indicating $\eta_{c}$ and $\eta_{c}^{\prime}$ \it resonances (left). A summary of the theoretical predictions and new experimental measurements of $\Delta M_{hf}(2S)$ \it (right).   
\label{etacp} }
\end{figure}
\par
The radial excitation of the charmonium spin singlet ground state, 
$\eta_{c}^{\prime}(2^{1}S_{0})$, is known to be bound. It is 
important to identify it because it can shed light on the nature of the 
spin-spin hyperfine interaction between a quark and antiquark. The hyperfine
interaction produces the splitting between the spin-singlet and spin-triplet 
states. For the charmonium $1S$ states splitting ($M(J/\psi)-M(\eta_{c})$) 
is known to be $\Delta M_{hf}(1S)=117 \pm 2$ MeV\cite{pdg}. It is important 
to know hyperfine splitting for the $2S$ states, because these states 
increasingly sample the confinement part of the $q\bar{q}$ potential.
\par
Crystal Ball has claimed observation of $\eta_{c}^{\prime}$ in an earlier 
measurement with $M(\eta_{c}^{\prime})=3594 \pm 5$ MeV in the $\psi^{\prime}$ 
inclusive photon spectrum\cite{crysballecp}. CLEO, with similar sensitivity, 
does not confirm the Crystal Ball observation\cite{skwarn}. 
In 2002 Belle announced the evidence for $\eta_{c}^{\prime}$ in two different 
measurements: in $B \to (\eta_{c}^{\prime})K \to (K_{s}K^{\pm}\pi^{\mp})K$ 
channel with $M(\eta_{c}^{\prime})=3654 \pm 6 \pm 8$ MeV\cite{belle1} and in 
double charmonium production $e^{+}e^{-} \to J/\psi \eta_{c}^{\prime}$ with 
$M(\eta_{c}^{\prime})=3622 \pm 12$ MeV\cite{belle2}. This was followed by 
CLEO\cite{cleoecp} (fig.\ref{etacp}) and BaBar\cite{babarecp} observations of 
$\eta_{c}^{\prime}$ in two-photon fusion processes with the results: \\
\hspace*{2.cm} CLEO \hspace*{5.5cm} BaBar \\
$M(\eta_{c}^{\prime})=3642.9 \pm 3.1 \pm 1.5$ (MeV), \hfill $M(\eta_{c}^{
\prime})=3630.8 \pm 3.4 \pm 1.0$ (MeV), \\
$\Gamma(\eta_{c}^{\prime})<31$ MeV ($90\%$ CL), \hfill $\Gamma(\eta_{c}^{
\prime})=17.0 \pm 8.3 \pm 2.5$ (MeV). \\
$\Gamma_{\gamma \gamma}(\eta_{c}^{\prime})=1.3 \pm 0.6$ (keV)\footnote{
Assuming that the branching fractions for $\eta_{c}$ and $\eta_{c}^{\prime}$
decays to $K_{s}K\pi$ are equal and using $\Gamma_{\gamma \gamma}(\eta_{c})=
7.4 \pm 0.4 \pm 0.5 \pm 2.3(br)$ (keV)\cite{cleoecp}.}.
\par
The world average of the $\eta_{c}^{\prime}$ mass value (fig.\ref{etacp}) is
$M(\eta_{c}^{\prime})=3637.4 \pm 4.4$ (MeV) and corresponds to hyperfine
mass splitting $\Delta M_{hf}(2S)=M(\psi^{\prime})-M(\eta_{c}^{\prime})=
48.6 \pm 4.4$ (MeV). This is a factor 2.4 smaller than
$\Delta M_{hf}(1S)$ and is not predicted by the potential model calculations
(fig.\ref{etacp}). This result should lead to a new insight into coupled
channel effects and the spin-spin contribution of the confinement part of 
$q\bar{q}$ potential.
\subsection{Two Body Hadronic $\psi(2S)$ Decays}
\begin{figure}[t]
\vspace{5.0cm}
\includegraphics{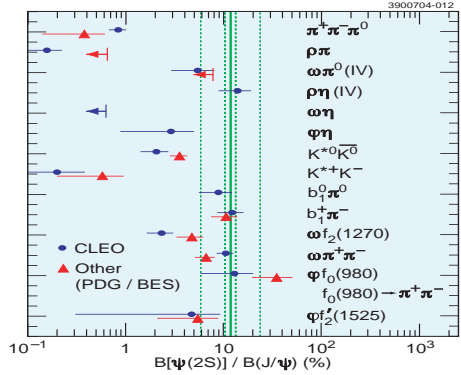}
\caption{\it The ratio of branching fractions of 
$\psi^{\prime}$ \it and $J/\psi$ decays to different final hadronic states. Vertical solid line indicates the pQCD expectation.
\label{psiphh} }
\end{figure}

According to pQCD, because both $^{3}S_{1} \to \gamma \to e^{+}e^{-}$
and $^{3}S_{1} \to ggg \to hadrons$ decays are proportional to 
$\mid \psi(0) \mid ^{2}$, the ratio 
\begin{equation}
Q_{h} \approx {{\cal B}(\psi(2S) \to h) \over {\cal B}(J/\psi \to h)} \approx
{{\cal B}(\psi(2S) \to e^{+}e^{-}) \over {\cal B}(J/\psi \to e^{+}e^{-})}
\approx (13 \pm 1) \%.
\label{rhopi}
\end{equation}
It was noted many years ago that the vector-pseudoscalar (VP) decay to 
$\rho \pi$ strongly violates the expectation of equation \ref{rhopi}. 
This problem is known  as the ``$\rho - \pi$'' puzzle and has received great 
theoretical attention. BES has recently measured vector-tensor (VT) 
($\omega f_{2}$, $\rho a_{2}$, $K^{*} \bar{K^{*}_{2}}$, 
$\phi f_{2}^{\prime}$) decays of $\psi^{\prime}$ with a data 
sample of $14 \times 10^{6}$ $\psi^{\prime}$ events\cite{beshh}. 
CLEO has measured $\psi^{\prime}$ decays to VP final states ($\rho \pi$, 
$\omega \pi$, $\rho \eta$, $K^{*0}\bar{K^{0}}$) and to $\pi^{+} \pi^{-} 
\pi^{0}$ with a data sample of $3 \times 10^{6}$ $\psi^{\prime}$ events. 
\cite{cleohh}. The results are summarized in fig.\ref{psiphh}.
\par
The experimental status of the ``$\rho - \pi$'' puzzle, based on the new 
measurements, can be summarized as follows: \\
- For VP final states, decays through three gluons are severely suppressed
with respect to the 13$\%$ rule and the corresponding isospin violating 
channels ($\omega \pi$, $\rho \eta$) are not; \\
- VT decay modes are suppressed by a factor of 3-5 compared to the 13$\%$ 
rule; \\
- Axial-pseudoscalar decay modes do not appear to be suppressed. 
\subsection{Radiative Transitions from $\psi(2S)$}
The measurements of radiative E1 electric dipole transitions ($\Delta L=1$, 
$\Delta S=0$) from $\psi(2S)$ were mainly done in 1980s by the Crystal
Ball\cite{crysball}. The latest improvements of these transition measurements
come from CLEO with a $\psi(2S)$ data sample comparable to the 
Crystal Ball sample. The preliminary CLEO results from the $\psi(2S)$ 
inclusive photon spectrum are\cite{skwarn}: 
${\cal B}(\psi(2S) \to \gamma \chi_{cJ}) = [9.75 \pm 0.14 \pm 1.17,9.64 
\pm 0.11 \pm 0.69,9.83 \pm 0.13 \pm 0.87]\%$ for $J=[2,1,0]$, 
respectively and for the ``hindered'' M1 transition: ${\cal B}(\psi(2S)
 \to \gamma \eta_{c})=(0.278 \pm 0.033 \pm 0.049)\%$.
\par
BES has measured the following branching fractions, using $\gamma \gamma
J/\psi$ events, from a sample of $14 \times 10^{6}$ $\psi(2S)$ decays:
${\cal B}(\psi(2S) \to \gamma \chi_{c1} \to \gamma \gamma 
J/\psi) = (2.81 \pm 0.05 \pm 0.23)\%$, ${\cal B}(\psi(2S) \to \gamma 
\chi_{c2} \to \gamma \gamma J/\psi) = (1.62 \pm 0.04 \pm 0.12)\%$, 
${\cal B}(\psi(2S) \to\pi^{0} J/\psi) = (1.43 \pm 0.14 \pm 0.12) \times 
10^{-3}$, ${\cal B}(\psi(2S) \to\eta J/\psi) = (2.98 \pm 0.09 \pm 
0.23)\%$\cite{bescascade}. A two photon cascade measurements from the 
CLEO data should be forthcoming soon.
\section{New in Upsilon Spectroscopy}
In this section new results are reviewed from the large data samples 
collected with the CLEO detector running at and in the vicinity 
of the $\Upsilon(1S)$, $\Upsilon(2S)$ and $\Upsilon(3S)$ resonances 
(about 20, 10 and 5 million events, respectively).
\subsection{First Observation of a $\Upsilon(1D)$ State}
$D$-wave states in charmonium are expected to be unbound and none, except 
the vector state at 3770 MeV, have ever been firmly identified. In 
bottomonium the 1$D$ and 2$D$ states are all expected to be bound but, 
until now, none had been identified. The $1^{3}D_{2}$ state has been 
identified with a significance of 10.2$\sigma$ at CLEO in the four photon
cascade (fig.\ref{spectras})\cite{dstate}: $\Upsilon(3S) \to \gamma 
\chi_{b}(2P)$, $\chi_{b}(2P) \to \gamma \Upsilon(1D)$, $\Upsilon(1D) \to 
\gamma \chi_{b}(1P)$, $\chi_{b}(1P) \to \gamma \Upsilon(1S)$, followed by 
the $\Upsilon(1S)$ annihilation into $e^{+}e^{-}$ or $\mu^{+}\mu^{-}$.
The measured mass $M(1^{3}D_{2}) = 10161.1 \pm 0.6 \pm 1.6$ (MeV) is in
agreement with both lattice and potential model calculations. The measured 
product branching ratio of the five decays is $(2.5 \pm 0.5 \pm 0.5) 
\times 10^{-5}$ and is also in agreement with theoretical estimates.
\subsection{${\cal B}_{\mu\mu}$ of the $\Upsilon$ States}
The total width ($\Gamma$) of the narrow $\Upsilon(1S,2S,3S)$ resonances 
produced in $e^{+}e^{-}$ interactions can not be measured directly because 
their natural width (25-50 keV) is much smaller than the energy resolution 
of an $e^{+}e^{-}$ collider (4-5 MeV). An indirect method of determining 
$\Gamma(\Upsilon(nS))$ is to combine the leptonic branching fraction 
(${\cal B}_{ll}$) with the leptonic decay width ($\Gamma_{ll}$), i.e.,
$\Gamma = \Gamma_{ll} / {\cal B}_{ll}$. Assuming lepton universality,
$\Gamma_{ll}$ can be replaced with $\Gamma_{ee}$ (CLEO plans to measure 
$\Gamma_{ee}$ with a few percent precision from scans of the resonant line 
shapes) and ${\cal B}_{ll}$ replaced with ${\cal B}_{\mu\mu}$. Therefore
the precise measurement of ${\cal B}_{\mu\mu}$ leads to a precise 
determination of $\Gamma(\Upsilon(nS))$.
\begin{figure}[t]
\vspace{5.0cm}
\includegraphics{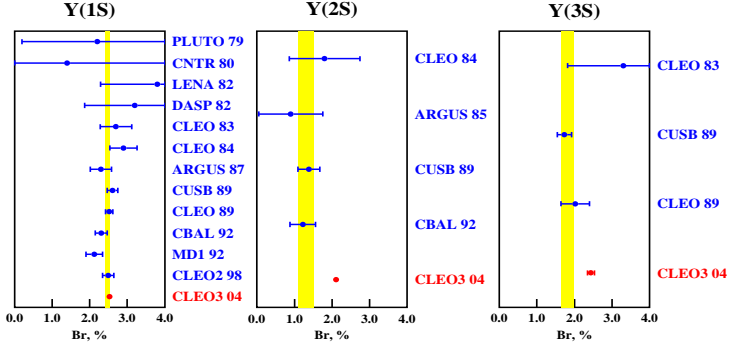}
\caption{\it Comparison of the new preliminary CLEO results of ${\cal B}(\Upsilon \to \mu^{+}\mu^{-})$ \it to other available measurements and the PDG average.   
\label{bmumu} }
\end{figure}
\par
CLEO has measured ${\cal B}_{\mu\mu}$ for the $\Upsilon(1S)$, 
$\Upsilon(2S)$ and $\Upsilon(3S)$ resonances by comparing muon and hadron 
yields at the peaks of resonances and the preliminary results are: 
${\cal B}_{\mu\mu}(\Upsilon(1S)) = (2.53 \pm 0.02 \pm 0.05)\%$, 
${\cal B}_{\mu\mu}(\Upsilon(2S)) = (2.11 \pm 0.03 \pm 0.05)\%$ and 
${\cal B}_{\mu\mu}(\Upsilon(3S)) = (2.44 \pm 0.07 \pm 0.05)\%$. 
The $\Upsilon(1S)$ result agrees with the PDG average\cite{pdg} but  
the $\Upsilon(2S,3S)$ results are significantly higher. They also imply 
narrower $\Gamma(\Upsilon(2S,3S))$. Results are shown in fig.\ref{bmumu}.
\begin{figure}[t]
\vspace{5.0cm}
\includegraphics{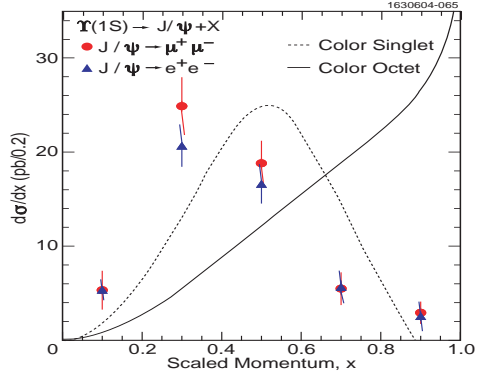}
\caption{\it $J/\psi$ momentum spectrum in $\Upsilon(1S) \to J/\psi X$ from the CLEO data. Theoretical expectations based on the color-octet and color-singlet models are shown with lines. 
\label{upspsi}}
\end{figure}
\subsection{$\Upsilon(1S)$ Decays to Charmonium Final States}
An explanation for the unexpected large charmonium production rates in 
$p\bar{p}$ collisions at the Tevatron was given by {\it color octet} 
models, where a single gluon fragments into a color octet $^{3}S_{1}$ 
$c\bar{c}$ pair which then evolves non-perturbatively into a color-singlet 
by emission of a soft gluon. {\it Color singlet} models produce final 
state $c\bar{c}$ mesons with two gluons. $\Upsilon(1S)$ decays are a good 
testing ground for the color octet and color singlet model predictions.
\par
CLEO has measured\cite{upspsi} the branching ratio 
{${\cal B}(\Upsilon(1S) \to J/\psi + X) = (6.4 \pm 0.4 \pm 0.6) \times
10^{-4}$} using $J/\psi \to \mu^{+}\mu^{-}$ and $J/\psi \to e^{+}e^{-}$ 
decays. Feed-down to $J/\psi$ from other charmonium states, e.g., 
$\psi^{\prime}$, $\chi_{cJ}$, is included. The color octet\cite{coloct} and 
color singlet\cite{colsing} model predictions of the branching fraction 
($6.2 \times 10^{-4}$ and $5.9 \times 10^{-4}$, respectively) are both
in agreement with the above result. However, the continuum subtracted $J/\psi$
momentum spectrum (fig.\ref{upspsi}) is in contradiction with the present
color octet model prediction.
\subsection{Neutral Dipion Transitions of $\Upsilon(3S)$ to $\Upsilon(1S)$ 
and $\Upsilon(2S)$}
Precise measurements of the dipion transition branching ratios for 
$\Upsilon(3S) \to \Upsilon(2S,1S)$ and dipion invariant mass spectra 
provide an experimental testing ground for many theoretical 
calculations\cite{dipion}, isospin conservation validation in charged and 
neutral dipion transition modes, and the deviation of dipion invariant mass 
from the phase space description. 
\par
CLEO has measured the following preliminary branching ratios: \\
\hspace*{2.cm} ${\cal B}(\Upsilon(3S) \to \pi^{0}\pi^{0} \Upsilon(2S)) = 
2.02 \pm 0.18 \pm 0.38$ ($\%$), \\ 
\hspace*{2.cm} ${\cal B}(\Upsilon(3S) \to \pi^{0}\pi^{0} \Upsilon(1S)) = 
1.88 \pm 0.08 \pm 0.31$ ($\%$). \\
The $\pi^{0}\pi^{0}$ effective mass spectrum from $\Upsilon(3S) \to 
\pi^{0}\pi^{0} \Upsilon(2S)$ has the shape consistent with several 
theoretical predictions. $\Upsilon(3S) \to \pi^{0}\pi^{0} \Upsilon(1S)$ 
was found to have a double peaked shape, also observed in the charged pion 
transitions\cite{dipion}. 
\section{New Narrow State $X(3872)$}
Belle recently observed a narrow state, $X(3872)$, in $B^{\pm} \to K^{\pm}X$,
$X \to \pi^{+}\pi^{-} J/\psi$, $J/\psi \to l^{+}l^{-}$, measuring $M(X) =
3872.0 \pm 0.6 \pm 0.5$ (MeV) and $\Gamma < 2.3$ MeV 
(90$\%$ CL)\cite{x3872belle}. CDF\cite{x3872cdf} and D0\cite{x3872d0} in $p\bar{p} \to X(3872)+...$, $X \to \pi^{+}\pi^{-} J/\psi$ and 
BaBar\cite{x3872babar}, in the same channel as Belle, confirmed this 
observation with $M(X) = [3871.3 \pm 0.7 \pm 0.4,3871.8 \pm 3.1 \pm 3.0,
3873.4 \pm 1.4]$ MeV, respectively.
\par
Many theoretical papers exist interpreting the $X(3872)$ state as: - a 
conventional charmonium state; - a $D\bar{D^{*}}$ molecule; - an 
exotic state. Identification of the quantum numbers is important 
to understand the structure of the state.
\par
\begin{figure}[t]
\vspace{5.0cm}
\includegraphics{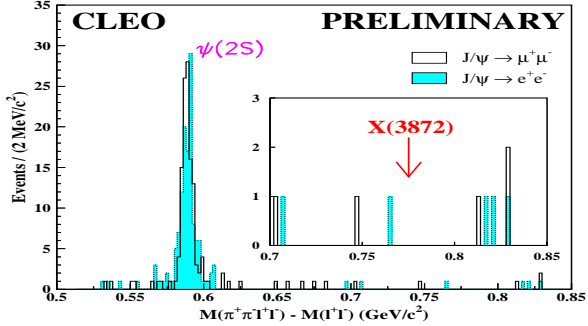}
\caption{\it Observed distribution of counts as a function of the effective mass difference $M(\pi^{+}\pi^{-}l^{+}l^{-})-M(l^{+}l^{-})$ \it from the CLEO data. The arrow indicates expected location of the $X(3872)$ signal.
\label{x3872}}
\end{figure}

CLEO has searched for $X(3872)$ with $\sim$15 $fb^{-1}$ of CLEO III data in
untagged $\gamma \gamma$ fusion production, where the state can be produced
if it has $J^{PC} = 0^{\pm +},2^{\pm +},...$, and initial state radiation 
(ISR) production, where the state can be produced if it has $J^{PC}=1^{--}$. 
The exclusive channels $X \to \pi^{+}\pi^{-} J/\psi$, $J/\psi \to l^{+}l^{-}$ 
were analyzed. No signals were found and the  following preliminary upper 
limits were set (fig.\ref{x3872}): 
\par
\begin{center}
$(2J+1)\Gamma_{\gamma \gamma}{\cal B}(X \to \pi^{+}\pi^{-} J/\psi) < 16.7$ eV 
(90$\%$ CL) in $\gamma \gamma$ fusion, \\
$\Gamma_{ee}{\cal B}(X \to \pi^{+}\pi^{-} J/\psi) < 6.8$ eV (90$\%$ CL) in ISR.
\end{center}
Systematic errors are included in the upper limits. 
\section{Summary}
Heavy quarkonium physics is an active field. Large data samples are being
collected and analyzed for quarkonia in $e^{+}e^{-}$ annihilation by 
BES-II ($c\bar{c}$), CLEO III ($b\bar{b})$, CLEOc ($c\bar{c}$). 
\par
Many new important experimental observations and measurements are available
and many others are expected. 
\par
Progress is being made in NRQCD and Lattice QCD calculations. Hopefully
many unresolved puzzles will be resolved soon.
\section{Acknowledgements}
I would like to thank the many collaborators on CLEO for providing analysis
results and discussions and also colleagues from BaBar, Belle and BES. \\
This work was supported by the U.S. Department of Energy.
\end{document}